\documentclass[manuscript]{aastex}

\usepackage{amsmath}
\usepackage{amssymb}
\usepackage{longtable}
\usepackage{array}
\usepackage{rotating}
\usepackage{multirow}
\usepackage{makecell}
\usepackage[graphicx]{realboxes}

\slugcomment{Not to appear in Nonlearned J., 45.}

\shorttitle{Chinese Sunspot Drawing and Its Digitization-(I) Parameter Archives}
\shortauthors{Lin et al.}

\begin{document}


\title{Chinese Sunspot Drawing and Its Digitization-(I) Parameter Archives}

\author{G.H. Lin\altaffilmark{1,2}, X.F. Wang\altaffilmark{1,2},  S. Li\altaffilmark{1,2},  X. Yang\altaffilmark{1,2}, G.F. Zhu\altaffilmark{1,2}, Y.Y. Deng\altaffilmark{1,2},  H.S. Ji\altaffilmark{3},  T.H. Zhou\altaffilmark{3}, L.N. Sun\altaffilmark{4}, Y.L. Feng\altaffilmark{5}, Z.Z. Liug\altaffilmark{5}, J.P. Taog\altaffilmark{5}, M.X. Beng\altaffilmark{5}, J. Lin\altaffilmark{5}, M.D. Ding\altaffilmark{6,7},  Z. Li\altaffilmark{6,7},  S. Zhengg\altaffilmark{8}, S.G. Zen\altaffilmark{8}, H.L. He\altaffilmark{8}, X.Y. Zeng\altaffilmark{8}, Y. Shu\altaffilmark{8},  X.B. Sun\altaffilmark{8}}

\affil{
$^{1}$Key Laboratory of Solar Activity, Datun Rd. 20A, Chaoyang District, Beijing, 100101, P. R. China\\
$^{2}$National Astronomical Observatories, CAS, Datun Rd. 20A, Chaoyang District, Beijing, 100101, P.R. China\\
$^{3}$Purple Mountain Observatory, CAS, 2 Beijing Xi Road, Nanjin, Jiangsu, 210008, P. R. China\\
$^{4}$Qingdao observatory,Purple Mountain Observatory, CAS\\
$^{5}$Yunnan Astronomical Observatoriy,CAS,396 Yanfangwang,Guandu District, Kunming, Yunnan,650216, P. R. China\\$^{6}$School of Astronomy \& Space Science, Nanjing university, 22 Hankou Road, Gulou District, Nanjng, Jiangsu, 210093, P. R. China\\
$^{7}$Key Laboratory for Modern Astronomy and Astrophysics, Nanjing University,Ministry of Education, Nanjing 210023, P. R. China\\
$^{8}$College of Science, China Three Gorges University, Yichang 443002, P. R. China\\}
\email{lgh@nao.cas.cn}
\email{lius@nao.cas.cn}


\begin{abstract}
Based on the Chinese historical sunspots drawings, a data set consisting of the scanned images and all their digitized parameters from 1925 to 2015 have been constructed. In this paper, we briefly describe the developmental history of sunspots drawings in China. This paper describes the preliminary processing processes that strat from the initial data (inputing to the scanning equipment) to the parameters extraction, and finally summarizes the general features of this dataset. It is the first systematic project in Chinese solar-physics community that the historical observation of sunspots drawings were digitized. Our data set fills in an almost ninety years historical gap, which span 60 degrees from east to west and 50 degrees from north to south and have no continuous and detailed digital sunspot observation information. As a complementary to other sunspots observation in the world, our dataset provided abundant information to the long term solar cycles solar activity research.

\end{abstract}


\section{Introduction}
The sunspot observations in the past provide the most direct data resource for the changes in solar activity. The conventional sunspot observations not only make to realize the variation of sunspots themselves, but also reveal some hints to understand other solar activity phenomena such as solar magnetic field, solar rotation, white-light flares \textit{etc.}
The formation of sunspots leads to the formation of active regions, which are the strongest magnetic fields in solar photosphere and are formed through the magnetic flux emergence from under the photosphere. Due to the magnetic field, the active region containing sunspots forms the main location where violent eruptions such as solar flares can occur (\citeauthor{2015AcASn..56..528L}, \citeyear{2015AcASn..56..528L}).
The intensive solar activity could have an impact on the Earth's magnetosphere and ionosphere, e.g. the telecommunication could be seriously hindered or even suddenly interrupted for a while, which will cause serious threat to the high technology system safety of aircraft, ships, and satellites, as well as telecommunication, facsimile and so on.

The long period temporal and spatial distribution evolution of sunspots reflects the process of magnetic dynamo from an interior of the sun. The research of solar dynamo problem relies heavily on long-term historical data accumulated by different human being. Therefore, it is significant to improve the time zone coverage of the observations and the accuracy of the extracted information from historical observing data, for understanding and predicting the long-term solar behavior, and also for studying the Sun influence on the solar--terrestrial space environment and our human activities (\citeauthor{2014AcASn..55..137T}, \citeyear{2014AcASn..55..137T};
\citeauthor{2015ChAA..39...45T}, \citeyear{2015ChAA..39...45T}).

The historical sunspot observations in the world are widely spread over different longitudes of the globe. The major sunspot information databases are given in Table~\ref{tab1}. In addition to these sunspot information database given in Table~\ref{tab1}, the World Data Center-SILSO(\citeauthor{2004SoPh..224..113V}, \citeyear{2004SoPh..224..113V}; \citeauthor{2014SSRv..186...35C}, \citeyear{2014SSRv..186...35C}, http://www.sidc.be/silso/) presents the production of total sunspot number (daily total sunspot number [1/1/1818 - now], Monthly mean total sunspot number [1/1749 - now], Yearly mean total sunspot number [1700 - now]) and the number of sunspot groups from 1610 to 2010. 
In this paper, our Chinese sunspot database is presented to fill the longitudinal gap in long term observations of sunspots from 1925 up to now to a great extent.

\begin{table}[h]
\centering
\caption{Current Major Sunspot Information Databases in the World}\label{tab1}
\begin{tabular}{|l|l|l|}
\hline
A catalogue of sunspot observations & 165 BC- &\citeauthor{1987AAS...70...83W}, (\citeyear{1987AAS...70...83W})\\
covering the period 165 BC to AD 1684 & AD 1684&\\\hline
University of Extremadura & 1610-2010 &http://haso.unex.es/haso/\\
(UE, Spanish)& &\\\hline
Greenwich Photoheliographic Results &1872-1976&\citeauthor{1984ApJ...283..373H}, (\citeyear{1984ApJ...283..373H})\\
(GPR, Britain) && \citeauthor{1991SoPh..136..251H}, (\citeyear{1991SoPh..136..251H})\\
& & \citeauthor{1993SoPh..146...27S}, (\citeyear{1993SoPh..146...27S})\\
&&\citeauthor{2013SoPh..288..141W}, (\citeyear{2013SoPh..288..141W})\\\hline
Kodaikanal Solar Observatory & 1906-1987 &\citeauthor{1984ApJ...283..373H}, (\citeyear{1984ApJ...283..373H})\\
(KK, India)& & \citeauthor{1991SoPh..136..251H}, (\citeyear{1991SoPh..136..251H})\\
& & \citeauthor{1993SoPh..146...27S}, (\citeyear{1993SoPh..146...27S})\\
& & \citeauthor{2013AA...550A..19R}, (\citeyear{2013AA...550A..19R})\\\hline
Mount Wilson Observatory	&1917-1985	&\citeauthor{1984ApJ...283..373H}, (\citeyear{1984ApJ...283..373H}) \\
(MW, USA)	&	&\citeauthor{1991SoPh..136..251H}, (\citeyear{1991SoPh..136..251H}) \\
& & \citeauthor{1993SoPh..146...27S}, (\citeyear{1993SoPh..146...27S})\\
&&\citeauthor{2004AGUFMSH52A..03U}, (\citeyear{2004AGUFMSH52A..03U})\\\hline
Debrecen Photoheliographic Data &	1974-now &	\citeauthor{2001MNRAS.323..223B}, (\citeyear{2001MNRAS.323..223B})\\
(DPD, Hungarian)&	 &\citeauthor{2015IAUGA..2257669B}, (\citeyear{2015IAUGA..2257669B})	\\\hline

\end{tabular}

\end{table}

Ancient China attached the importance to astronomical records because the Emperor's divine rights were explained to be granted by Heaven. Actually, China seems to be the first country in the world to record sunspots (traceable around 364 B.C., \citeauthor{2011nrichL}, \citeyear{2011nrichL}) and has long-term text records of sunspots, however, scientific quantification was not implemented. The telescope was introduced in the ancient China between the Ming and Qin Dynasties in 1622 by the preacher Johann Adam Schall Von Bell, which played an important role in the brutal palace battle of calendar between the old one and the new one through the observation. From solar observations, the people at that time recognized that the Sunspots "drift from east to west, 14 days along the diameter, the larger one reducing the Sun's luminosity" (\citeauthor{2008g}, \citeyear{2008g}). China may have a lot of sunspots drawings through telescopes, but at present, only the largest relatively comprehensive and systematic observations are gathered in our data set. This data set consists of observations from six stations, but the following three stations contribute the vast majority of data to our archives.

\begin{figure}
   \centerline{\includegraphics[width=1\textwidth,clip=]{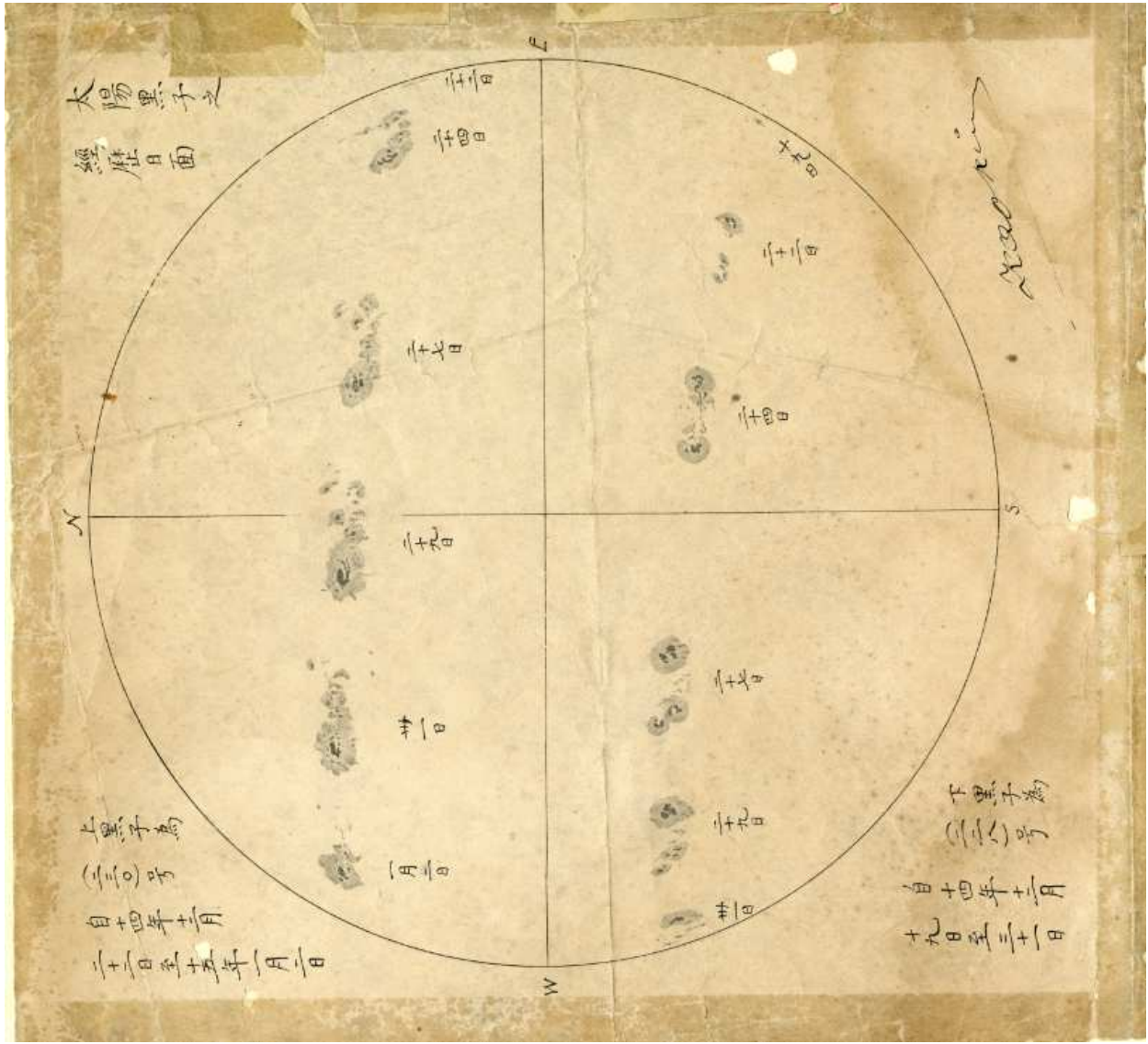}}
   \caption{Early original picture of sunspot drawing made by Gao Pingzi at Qingdao observatory. This is the combined sunspot evolution images that observed on various date: for up sunspot the date is from 22 Dec 1925 to 2 Jan 1926, while for the bottom sunspot the date is from 19 to 31  1925. For positions, they are labeled by N, S, W and E in the image.}\label{fsd}
\end{figure}

\emph{Qingdao Observing Station} (QDOS) is the earliest in modern China to study sunspots using telescope, but its sunspot drawings had a rather difficult history (see Table~\ref{qdot}). Its first observation in current archives was recorded by Gao Pingzi (1888--1970) on 1st May, 1925. An example of the same year observation is shown in Figure~\ref{fsd}. He used a 16~cm equatorial telescope left by German and installed a projector board behind the eyepiece. Then sunspots and plage were depicted by hand on a paper paved on this board, adjusting the solar diameter to 18.2 cm. One picture is drawn every day except when it was overcast.
\begin{table}[h]
\centering
\caption{Brief History of Qingdao Observatory Station after the Establishment}
\begin{tabular}{|l|l|}
\hline
1937-1945& Observation continued during the Japanese occupation. \\
        & After the war, part of data, the 16 cm objective lens\\
        &and photographic devices were plundered.\\\hline
1947 &Sunspots drawing resumed with a 4-inch lens and 14.4 cm solar \\
        &diameter on the projection board.\\\hline

1954 & Objective lens was updated to 15 cm. Focal length 2.2 m. \\
        & Projected solar diameter is the national uniform standard 17.4 cm.\\\hline
1978-1983&	Observatory was rescinded. Observation stopped.\\\hline
1983-1988&Supported by Chinese Academy of Sciences, observation was \\
        &resumed using a modified 20 cm aperture guide telescope with \\
        &a 3.5 m focal length and solar projection diameter 17.4 cm. \\\hline
1988-now &	32 cm refractive telescope equipped with sunspots\\
          &	 fine structure carmera. The penumbra fibril resolution reaches 0.7$^{\prime\prime}$.\\
       &The penumbra of sunspots in the drawings can be seen.\\\hline

\end{tabular}
\label{qdot}
\end{table}

\emph{Purple Mountain Astronomical Observatory} (PMO) carried out sunspot drawings with a 20~cm aperture refractive telescope made by Zeiss company 
in 1934. In 1937, PMO was forced to move away from Nanjing, but in the same year, the PMO staff set up an observatory at the top of Fenghuang Mountain in the vicinity of Kunming city. Sunspot observations were made with an 8~cm refractive telescope and relative sunspot numbers were published every half year. After the victory of the war against Japan, most of PMO staff moved back to Nanjing to rebuild the observatory that was in destruction. In 1954, all sunspot drawing data in China was collected and analyzed at PMO. Since then, regular observations of daily sunspot drawings in China started, and in 1957, all sunspot drawings were required to have solar disk image of 17.4 cm. 

\emph{Yunan Astronomical Observatory} (YNAO) was developed from PMO's station in Fenghuang Mountain after the war with Japan. YNAO's sunspots drawings tradition was inherited from PMO and its observation continuity make it best in our data set. So, YNAO contributes the largest part to our sunspots drawings archive. The sunspot drawings from YNAO were evaluated by international colleagues. Sunspots area data for each day were found to have the smallest random error in the world and they had good quality to supplement the global sunspot data, and also filled time zone gaps (\citeauthor{2001MNRAS.323..223B}, \citeyear{2001MNRAS.323..223B};
\citeauthor{2009JGRA..114.7104B}, \citeyear{2009JGRA..114.7104B}).

\textit{Chinese Solar-Geophyscal Data} (CSGD, printed journal) from 1971 to 2001, has published the  sunspots' daily relative numbers,  areas
and predicted smoothed numbers, which all came from the records of suspot drawings from Chinese observatories (\citeauthor{yan2018}, \citeyear{yan2018}). Parts of CSGD are published online by NGDC (National Geophysical Data Center), whose website is http://www.ngdc.noaa.gov/nndc.

In general, Chinese sunspot drawings contain almost complete physical information realized in that era. However, these materials were not easy to be completely preserved due to the difficulties in long history, humidity, decay, pests damage, relocations, \textit{etc.} In order to better utilize the scientific value of these data, it is necessary to extract information completely, accurately, and reliably from Chinese historical sunspot drawings. Also, make the data available through network sharing and keep the data for long term use by digitizing the sunpot drawings so that the information from the data is preserved.

With the support of National Basic Research Program from the Ministry of Science and technology of China, the digitization of Chinese historical sunspot drawings commenced from May 2014. By the end of April 2018, we finished the digitization of sunspots drawings from six Chinese observing stations and also completed their parametric extraction. In the process of digitization we go through these processe: the scanner selection, the scanning results evaluation of original image, the automatic extraction of handwritten parameters, the proof of extracting parameters by computer program, and manually checking through sampling. Here, we presented a part of statistical results from the entire analysis, and the details of process are given in the following six sections.

\section{Observation Data of Chinese Sunspot Drawing}
\begin{figure}
   \centerline{\includegraphics[width=1\textwidth,clip=]{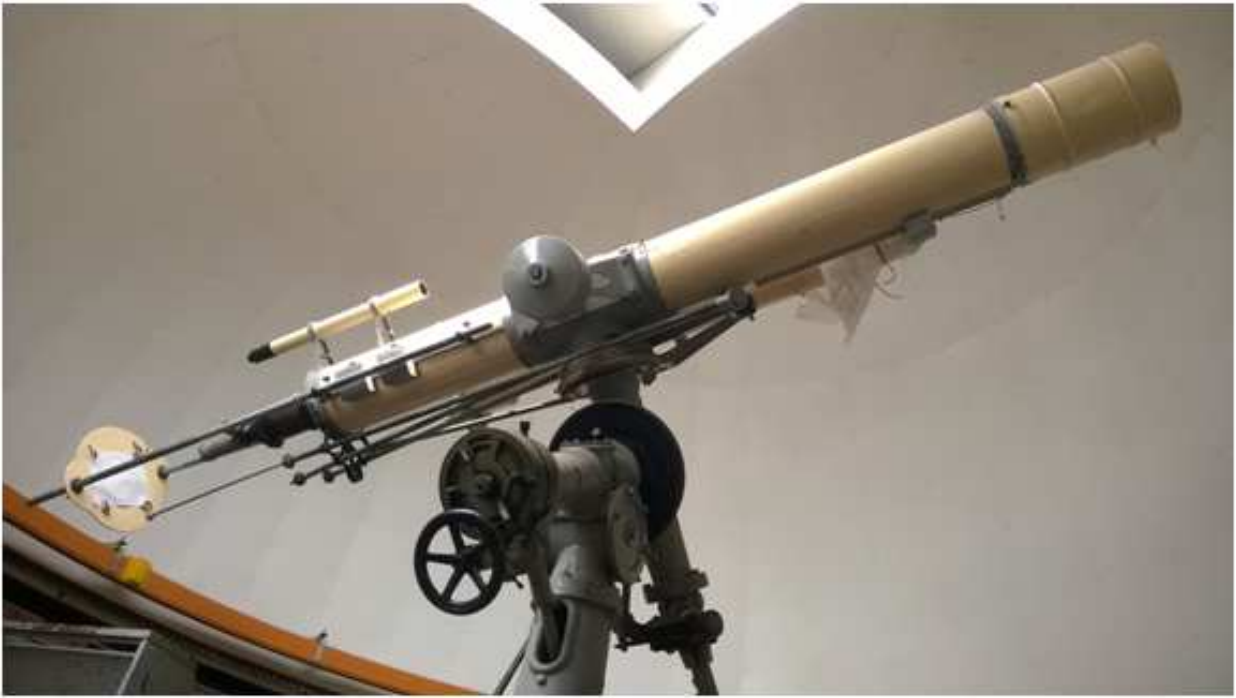}}
   \caption{PMO sunspot drawing telescope, it is an equatorial refractive telescope with the diameter and focal length of 20 cm and 350 cm. When it is observing, the radius of the sun's projection is 17.4cm. } \label{zt}
\end{figure}

\begin{figure}
   \centerline{\includegraphics[width=1\textwidth,clip=]{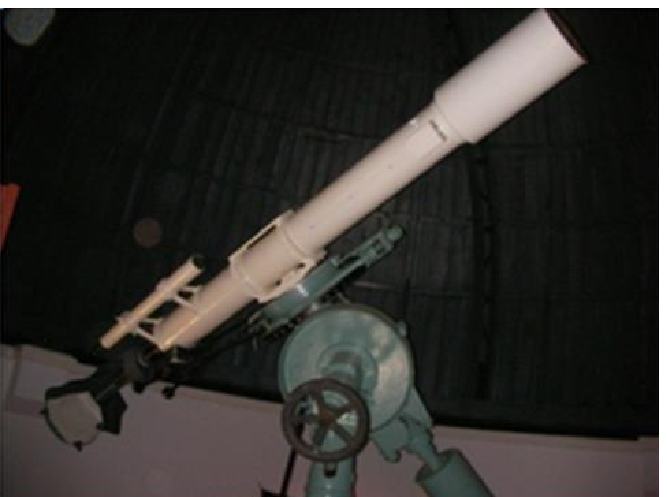}}
   \caption{YNAO sunspot drawing telescope, this is an equatorial refractive telescope with the diameter and focal length of the diameter and focal length are 12.7 cm and 195 cm. When it is observing, the radius of the sun's projection is 17.4cm.} \label{yt}
\end{figure}

Historically, there mainly exists six observing stations for sunspot drawings in China: PMO, YNAO, QDOS, \emph{Sheshan Observing Station} (SSOS), \emph{Beijing Planetarium} (BJP) and Nanjing University (NJU). 
Table~\ref{sdlisttab} lists Chinese sunspot drawing data, where the name, longitude and latitude of station, time range, number of images are recorded. The data continuity is best achieved by YANO, with a time span from 1957 to now (even if afterwards such traditional observations were no longer supported by operational funds), of 62 years. The second best, PMO with a time span of 57 years. At each individual station on sunny days, one sunspot image is drawn. In China, the first sunspot drawing was made at the QDOS in 1925. Unfortunately, the observations were soon interrupted by Japanese invasion of Nanjing in 1937 and sunspot drawing data of nearly two years were lost. Additionally, the observations were interrupted several times due to social turmoil of that time. Regular observations of daily sunspot drawings took place only after 1947. Since then, the observatory has been continuously performing observations. Table~\ref{sdlisttab} shows that there are different observation periods for data of QDOS and PMO. However, from the perspective of complementarity of observations, there exists continuous sunspot observation data in China from 1947 to now. The main parameters of sunspot drawing telescope in all observing stations are listed in Table \ref{equipments}. In this paper, we mainly discuss the data processing of sunspot drawings by PMO and YNAO as examples among six observing stations. 
 Figure \ref{zt} and \ref{yt} are sunspot drawing telescopes of PMO and YNAO, respectively.


\begin{table}[h]
\centering
\caption{Information About Station and Observation Years in China}
\begin{tabular}{|l|l|}
\hline
Observatory/Station (Abbreviation)&Observation Years\\\hline
Qingdao Observatory Station (QDOS)&1925\\
N$36^\circ 04^\prime$, E$120^\circ 19^\prime $&1947-1977\\
&1980\\
&1982-1989\\
&1991-1992\\
&1995-1996\\
&2000-2009\\
&2011-now\\\hline

Purple Mountain  Astronomical Observatory (PMO)&1954-1963\\
N$32^\circ 03^\prime$, E$118^\circ 49^\prime $&1965-1980\\
&1982\\
&1985-2011\\\hline

Yunnan Astronomical  Observatory (YNAO)&1957-2016\\
N$25^\circ 01^\prime$, E$102^\circ 47^\prime $&2018-now\\\hline

Sheshan Observatory  Station (SSOS)&1952-1964\\
N$31^\circ 06^\prime$, E$121^\circ 13^\prime $E&\\\hline

Beijing Planetarium (BJP)&1979-1982\\
N$39^\circ 54^\prime$, E$116^\circ 23^\prime $&1989-1999\\\hline
Nanjing University (NJU)&1986-2002\\
N$32^\circ 03^\prime$, E$118^\circ 51^\prime $&2004-2015\\\hline
\end{tabular}
\label{sdlisttab}
\end{table}

\begin{table}[h]
\centering
\caption{Main Parameters of Sunspot Telescope for the Most Recent Individual Stations.}
\begin{tabular}{|l|l|l|l|}
\hline
Observatory/Station&Diameter&Focal Length&Diameter of the sun\\\hline

QDOS&32 cm&350 cm&17.4 cm\\\hline

PMO&20 cm&350 cm&17.4 cm\\\hline
YNAO&12.7 cm&195cm&17.4 cm\\\hline
NJU&43 cm&217cm&20 cm\\\hline
\end{tabular}
\label{equipments}
\end{table}



\subsection{Processes of Sunspot Drawing}
Traditional projection method was used by PMO, YNAO, \textit{etc.} wherein they obtain sunspots drawing by projecting an enlarged image of the Sun onto a projection plate. At first the preprinted sunspot observation record paper is fixed on the projection plate, and the position of directions (east, west, south and north) is determined accurately. Then, slowly the telescope is moved to make sure that the solar projection always overlaps with the solar limb printed on the recording paper. Finally, the specific information of sunspots is accurately drawn with a pencil. For example, according to the projection image of sunspots on the projection plate, the penumbra of the sunspots is firstly drawn with a hard pencil, and then the umbra of the sunspots is traced with a soft pencil. The western sunspots are drawn before the eastern ones. The larger sunspots are drawn first and then the smaller ones. After the observations, other conventional observing information is recorded: the date of observation, Beijing time (standard time of 120 degrees east longitude), international standard time (UTC) and Carrington rotation number (on November 9, 1853, the moment, when prime meridian switch to center of solar disk, is defined as the beginning of the first solar rotation, from then the solar rotations are numbered, the start and end dates and numbers for each solar rotation can be found in astronomical almanacs). At the same time, through the astronomical almanacs , the related parameters at the observation time are calculated, such as P (the position angle between the geocentric north pole and the solar rotational north pole, in the Carrington coordinates of the solar surface, positive toward the east and negative toward the west), B$_{0}$ (heliographic latitude of the center of the solar disk observed at the universal time of zero on that day), L$_{0}$ (heliographic longitude of the center of the solar disk at zero universal time on the observation day), L (heliographic longitude of the center of solar disk at the time of the observation). These data can be used to obtain  heliographic coordinates, \textit{i.e.} to calculate the latitude and longitude of the sunspots. Also, the area of the sunspot group can also be measured: A special transparent glass plate is placed on the projection plate. Each square area in the glass plate is 1 square millimeter and the number of squares corresponding to each sunspot group is recorded (\citeauthor{2016AcASn..57..292L}, \citeyear{2016AcASn..57..292L}). The number of squares contained in a sunspot group can be converted into the area of sunspots by correlation calculation. Then, the sunspot group is numbered and other information such as the coordinates of the sunspot group and the types of the sunspot group are recorded. Finally, the number of sunspots groups, the number of sunspots, the Wolf number on the southern hemisphere, the northern hemisphere and the whole solar surface are calculated, respectively (\citeauthor{2012r}, \citeyear{2012r}).
\subsection{The Content of Sunspot Drawing}
\label{The Content of Sunspot Drawing}
\begin{figure}
   \centerline{\includegraphics[width=1\textwidth,clip=]{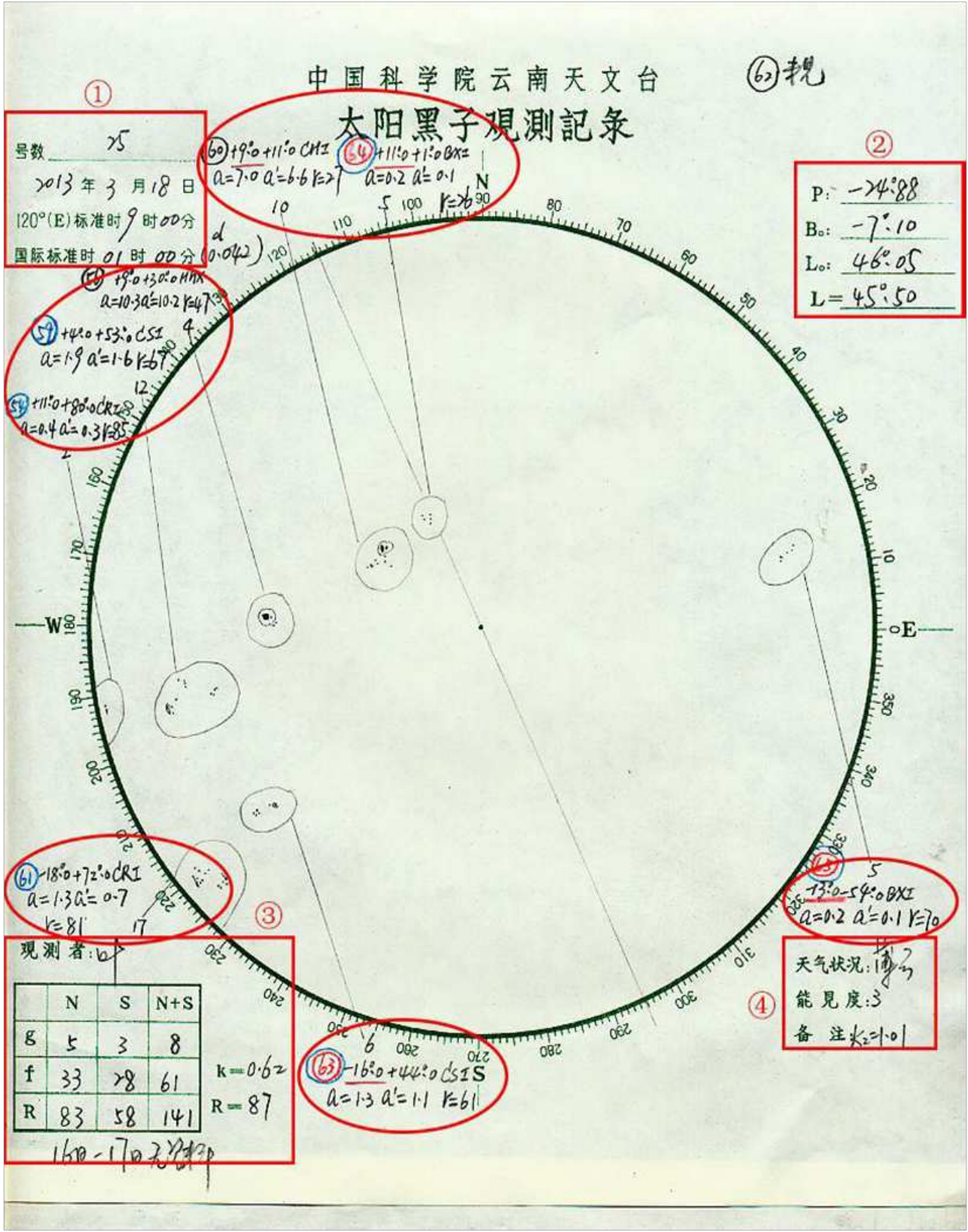}}
   \caption{An example of Sunspot drawing observed by YNAO obseved on 18 March 2013 01:00 UTC. Here printed part information is recorded in the rectangle region, while the handwriten part information relating specific and individual sunspot is recorded in the oval boxes. The orientation (N, S, W and E) and angle-scale (Degree) are marked on the circumference.} \label{sd1}
\end{figure}

An example of sunspot drawings is shown in Figure \ref{sd1}, in which the content of the rectangular and oval boxes marked with red color are described below.
The rectangular boxes are printed at fixed position on the observation paper.
Rectangular box 1 contains:\\
Observing day number within the year.\\ Date of observation.\\ Beijing time.\\Universal Time Coordinated (UTC).\\
Rectangular box 2 contains:\\
P angle: the position angle between the geocentric north pole and the solar rotational north pole measured eastward from geocentric north.\\
B$_{0}$ : heliographic latitude of the central point of the solar disk.\\
L$_{0}$ : heliographic longitude of the central point of the solar disk.\\
L: heliographic longitude of the center of solar disk at the time of the observation.\\
Rectangular box 3 contains various combinations:\\
g: number of sunspot group.\\
gN: number of sunspot groups in the northern hemisphere.\\
gS: number of sunspot groups in southern hemisphere.\\
gNS: total number of sunspot groups, gNS=gN+gS.\\
f: total number of sunspot.\\
fN: number of sunspots in northern hemisphere.\\
fS: number of sunspots in southern hemisphere.\\
fNS: total number of sunspots, fNS=fN+fS.\\
R: Wolf number.\\
RN: Wolf number in the northern hemisphere.\\
RS: wolf number in the southern hemisphere.\\
RNS: total Wolf number, RNS=RN+RS.\\
K: the normalized coefficient of sunspot relative number, that related to site instrument and observere and it varied with time. Here Rz published in Zurich was used as the standard, K = R$_z$/R$_y$ (R$_y$=RNS). After the stop of Zurich, the standard of international sunspot relative number R$_i$ is used as standard. \\
K$_{2}$: It is similar to K, but for normalized coefficient of sunspot area.\\
The observer's last name.\\
Rectangular box 4 contains:\\
Weather condition: \emph{e.g.} thin cloud.\\
Seeing:  \emph{e.g.} 3. Their values range from 1 to 5, in which 1 is the best.\\
The information of the oval boxes are the random handwritten region information. The contents are the description of sunspots on the surface, including:\\
Sunspot group number.\\
Longitude, latitude.\\
Structure types of sunspot: \emph{e.g.} CHI, BXI, here the McIntosh classification was used (\citeauthor{1990SoPh..125..251M}, \citeyear{1990SoPh..125..251M}).\\
$\alpha$: total area of the sunspot group, one in a million of area of solar disk.\\
$\alpha^{'}$: the area of the largest sunspot in the sunspot group, units of millionth of thesolar disk, referred to as the area of the largest sunspots.\\
r: linear distance between center of mass of the sunspot group and the center of the solar disk in units of mm.\\
For example in Figure~\ref{sd1} the oval handwritten area for the No. 63 sunspot group gives, latitude and longitude of -16.0$^\circ$ and + 44.0$^\circ$, sunspot group structure type CSI group area of $\alpha$ = 1.3, the area of the largest sunspot $\alpha^{'}$ = 1.1, linear distance between center of mass of sunspot group and the center of the solar disk r = 61.

\begin{figure}
   \centerline{\includegraphics[width=1\textwidth,clip=]{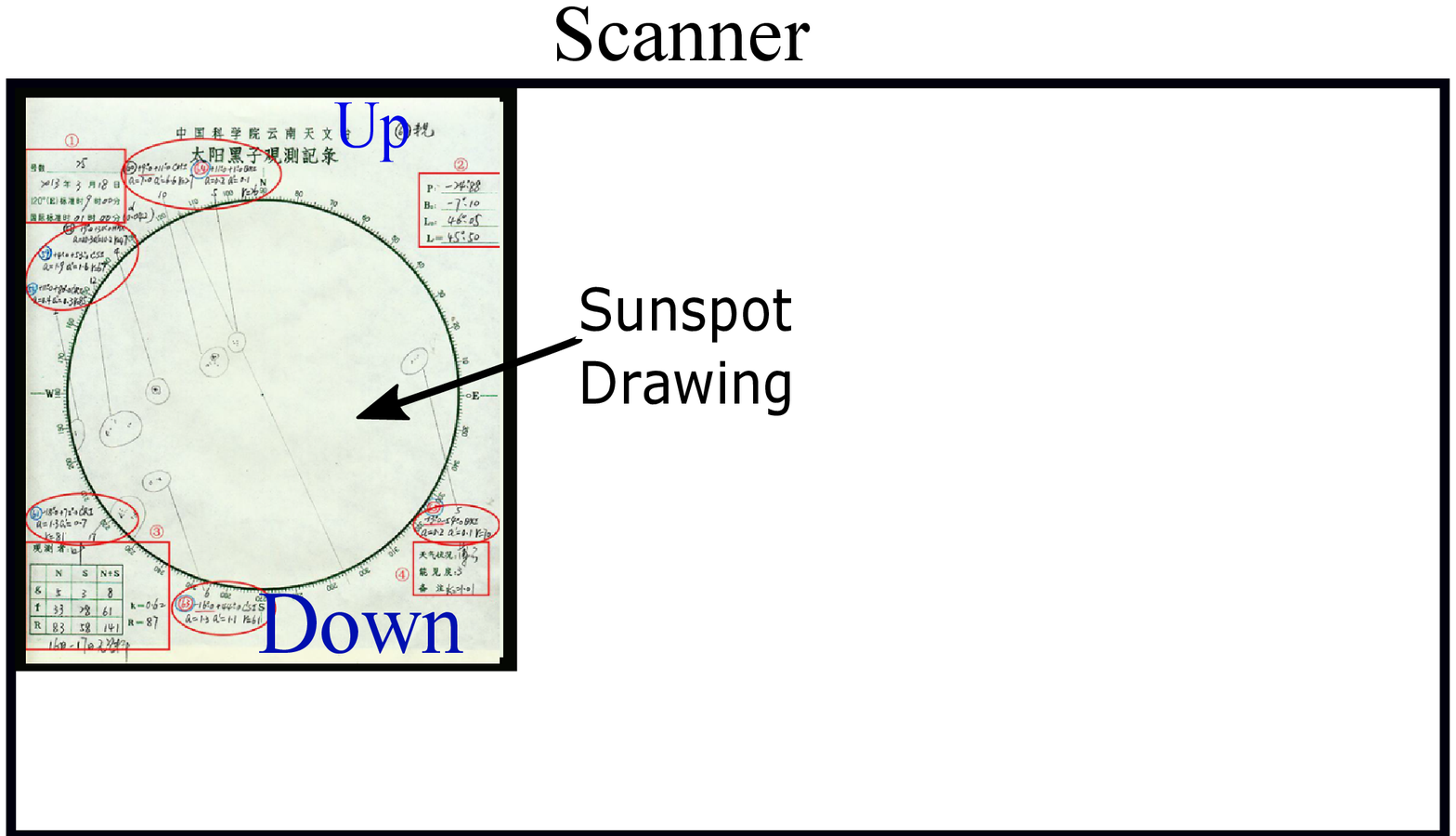}}
   \caption{The position of the sunspot drawing in the scanner. The upper and left edges of the image coincide strictly with the scanner, the north of the image is above the printer, while the west is stuck to the left. At the same time, try ot the best to keep the paper flat and close to the printer.} \label{scanner}
\end{figure}

\section{Digital Processing}
For the digital processing, we visited World Data Center in Brussels, Belgium and made collaboration with Dr. Frederic Clette who was converting Europe's historical sunspot drawings into digital files. Referring to their standard suggestions and many scanning trials, we established a few regulations for the guidance of ever-repeated scanning work.

\subsection{Scanner selection}
For the selection of the scanner, there are three basic requirements, 1) resolution: the description of the smallest spot on the sunspot drawing is about the size of the pencil tip (0.05mm). To clearly scan such a point, at least 2 pixels are needed, which requires the resolution of the scanner to be $>$1000 ppi (pixels per inch). 2) gray level: in order to effectively distinguish the umbra from the penumbra (due to the long storage time of the early drawings and light color), the scanner with the maximum gray level is required. 3) width: A3. After careful investigation, Zhongjing 1960XL scanner that met the above requirements was chosen.

\subsection{Scanning Rules}
The rules of scanning originates from two aspects: Visions and qualities. The sources of different time (year) and observers are selected and put into the scaner. The main operation is to adjust the brightness and contrast, for example, when the brightness of gray scale of image is too bright or too dark, then drag the brightness slider to change it. If the brightness is too high, the image will look white. It will be too dark if the brightness is too low. The brightness of the image should be moderate while dragging the brightness slider. For other parameters, the local changes can be made according to the similar adjustment method. The team members carried out visual inspection until everyone was satisfied, and then formulated the technical parameters and operational rules for the scanning of sunspots. Quality assessment was carried out by comparing the relative number of sunspots between digital and original data to find out uncover potential, and repeatedly verified, the preparation ends when requirements (relative error $<$ 5\%) are satisfied.
After the scanner is adjusted according to the above rules, the specific rules of scanning process is as follows: 1) prepare a piece of pre-scanned original sunspot drawing, check the condition of the folding, wrinkle and dust on the surface, and carefully clean the paper surface and flatten it. 2) Put the original image face down into the scanner, adjust the position, and align the scanned paper with the corner of the scanner. Make sure the placement position that the north--south direction of the observation paper is perpendicular to the scanning direction, as shown in Figure \ref{scanner}.
3) In order to save all the details of the original observation, the 24-bit BMP (Bit Map Picture) color image format was chosen to save the image, resulted in the size of single image about 90 MB. Although this data format is more than 10 times larger than the commonly used JPG (Joint Photo Graphic) format, the original observations can be truly recorded without compression distortion (Figure~\ref{sd1}), which allowed us to work with great precision to analyze each of the images.

\subsection{File naming and Storage}
Since the data was produced from multiple stations, the file name and storage directory of all observational data that were scanned are specified for the convenience of their database query. The file name is consistent with the basic rules of international astronomical naming, and the unified form is 

$<observatory>$$\_$$<sd>$$\_$$<year><month><date>$$\_$$<hour><minute>$$\_$$<observer>.bmp$

The observatory refers to the source of observational data, the abbreviation and corresponding full name of China observatory  can refer the first column of Table~\ref{sdlisttab},
Sd refers to the type of observational data which indicates sunspot drawing. The format of observation time is year, month, date, hour and minute (UTC). For example, for the sunspot drawing painted by YNAO observer whose last name is Ye at 5:40 on February 18, 1957, the file name after digitization is as follows:

$Ynao$$\_$$sd$$\_$$19570218$$\_$$0540$$\_$$ye.bmp$

After scanning, the file is stored on the server, and the folder is set as follows:
/1957/02/YNAO/
\section{Automated Extraction of Parameters}
\begin{figure}
   \centerline{\includegraphics[width=1\textwidth,clip=]{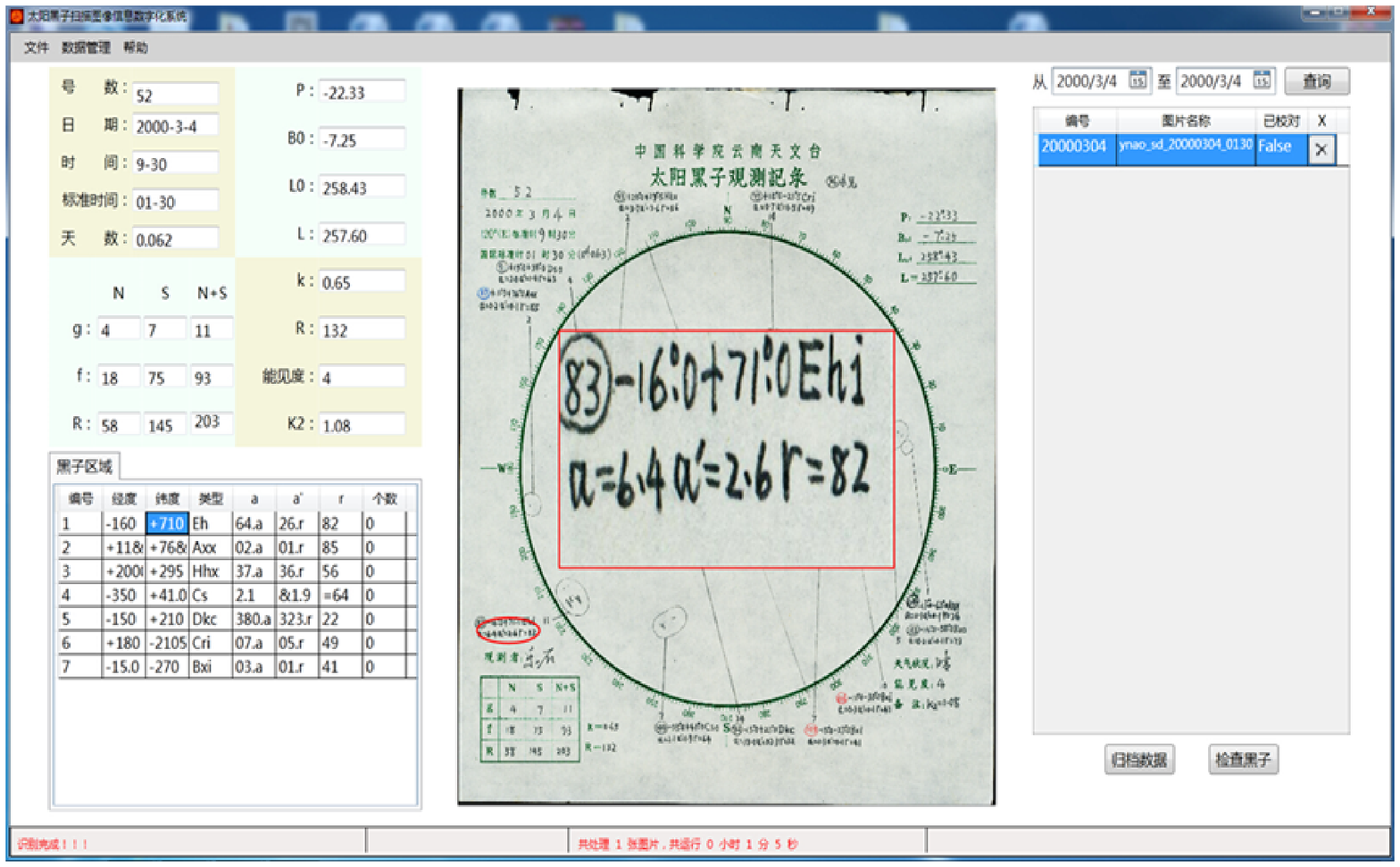}}
   \caption{Sunspot drawing identification software. The left parts show the identifiable information that use automatic recognition method; The middle part shows the original image and obseved information; While the right parts give the some needed operations such as archiving and proofreading.} \label{software}
\end{figure}

Based on the scanned image and convolutional neural network (CNN), an identification software is developed to recognize the sunspot drawing parameters automatically, which is further described in Section~\ref{The Content of Sunspot Drawing} of the automatic identification of the content in the rectangular (printed parts) and oval box (hand-written parts). Figure~\ref{software} shows the interface of identification software, wherein the content on the left is the recognition results shown by the middle image, which includes the whole content of rectangular and oval box. For a detailed introduction of this software, we refer the reader to a former paper (\citeauthor{2016NewA...45...54Z}, \citeyear{2016NewA...45...54Z}).
\begin{figure}
   \centerline{\includegraphics[width=1\textwidth,clip=]{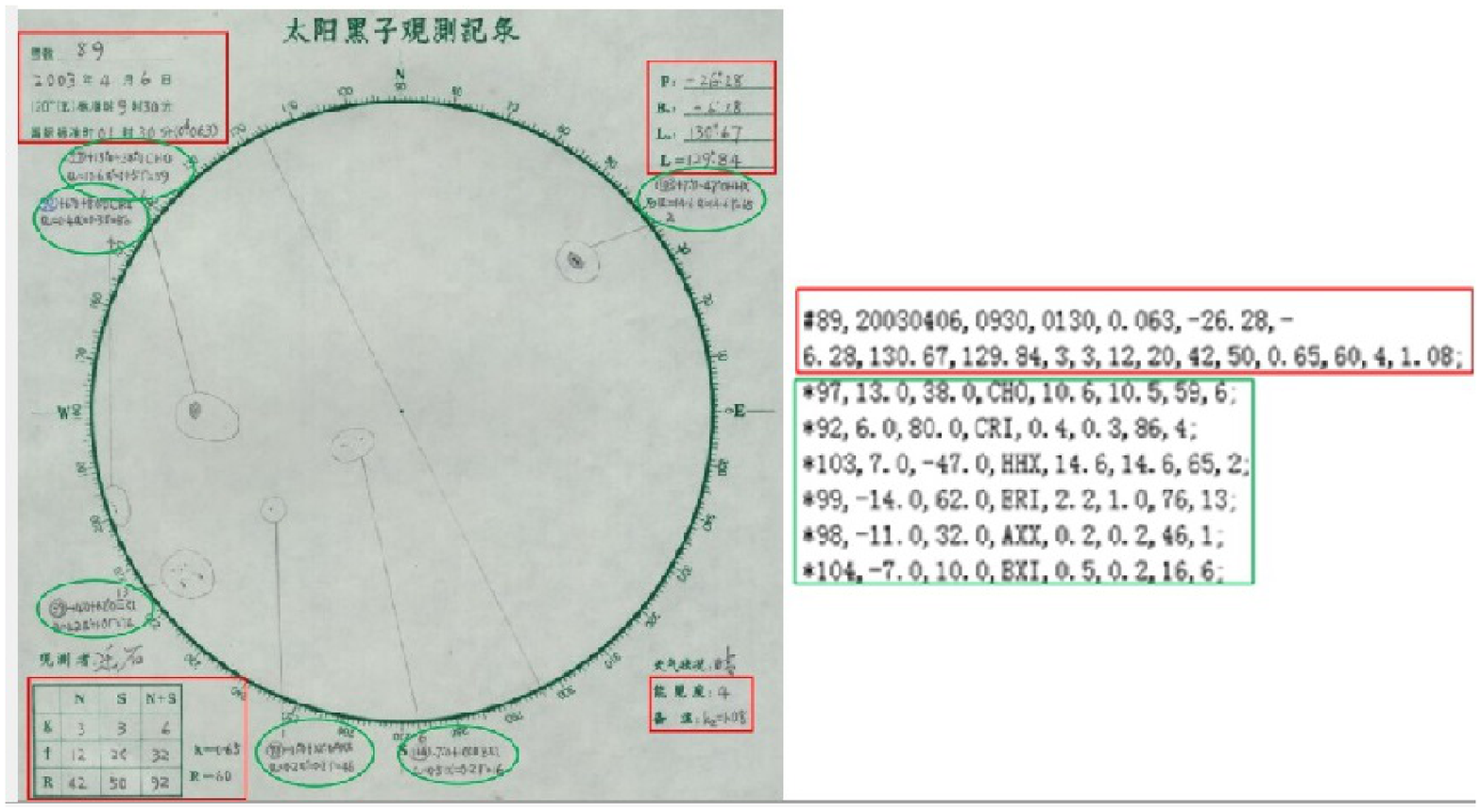}}
   \caption{TXT record format and its corresponding image content. The red part region labeled in TXT corresponds the printed information indicate by red rectangle in original image; While the green part region labeled in TXT is the information identified from the handwriten character in original sunspot drawings.} \label{softwareresult}
\end{figure}
\subsection{Parameter Record Format}
\label{Parameter Record Format}
The identified parameters are stored in the ".txt" document format (on the right of Figure~\ref{softwareresult}) and the".csv" table format (see Table~\ref{tabcsv1},\ref{tabcsv2}), which are available to users for further processing them as required.
In the TXT document format, the printed part of the data content begins with '\#', ',' is separator,
';' means end, '?' indicates that the information is unknown (or, due to the longer time the handwriting is blurred and cannot be recognized; or here is no original information), there is no space between the separator. The handwritten part of the data content begins with '$\ast$', ',' is separator, ';' is end, '?' means the information is unknown (the reason is the same as above), as shown in Figure~\ref{softwareresult} on the right. An example of records is shown in Figure~\ref{softwareresult}, wherein the information is labeled by red/green color boxes that were recorded as red/green rectangle labeled in documents, respectively.
\begin{table}[h]
\caption{CSV Table Printed Parts} 
\centering 
\setlength{\tabcolsep}{1.2mm}{
\Rotatebox{90}{%
\begin{tabular}{ccccccccccccccccccccccccc} 
\hline\hline 
Num.&Date & BJ$_{time}$ &UTC&Day&P&B$_{0}$&L$_{0}$&L&gN&gS&gNS&fN&fS&fNS&RN&RS&RNS&K&R&Vis.&K$_{2}$
\\ \hline 
286&	19841110&	850&	50&	0.035      &22.67&	3.37&	   276.24&	275.78&	2&	0&	2&	14&	0&	14&	34&	0&	34&	0.65&	22&	3&	0.99\\   
287&	19841111&	1020&	220&	0.097&	22.43	&    3.26&   263.08&	261.77&	2&   1&	3&	14&	3&	17&	34&	13&	47&	0.65&	31&	3&	0.99\\
288&	19841112&	900&	100&	0.042&	22.18&	3.14&	  249.87&	249.32&	1&	1&	2&	10&	1&	11&	20&	11&	31&	0.65&	20&	3&	1.08\\
289&	19841113&	850&	50&	0.035&	21.92&	3.03&   236.69&	236.23&	1&	0&	1&	8&	0&	8&	18&	0&	18&	0.65&	12&	3&	1.08\\
290&	19841114&	840&	40&	0.028&	21.65&	2.91&   223.5&	223.13&	1&	0&	1&	4&	0&	4&	14&	0&	14&	0.65&	9&	2&	1.08\\
291&	19841115&	900&	100&	0.042&	21.38&	2.8&	  210.32&	209.77&	0&	1&	1&	0&	3&	3&	0&	13&	13&	0.65&	8&	2&	1.08\\
292&	19841116&	940&	140&	0.07	&    21.10&     2.68&	  197.14&	196.22&	0&	1&	1&	0&	2&	2&	0&	12&	12&	0.65&	8&	3&	1.08\\
293&	19841117&	905&	105&	0.045	&    20.81&	2.56&	  183.95&	183.35&	0&	1&	1&	0&	2&	2&	0&	12&	12&	0.65&	8&	2&	1.08\\
294&	19841118&	835&	35&	0.024&	20.52&	2.44&	  170.77&	170.44&	1&	0&	1&	4&	0&	4&	14&	0&	14&	0.65&	9&	2&	1.08\\
295&	19841119&	830&	30&	0.021&	20.22&	2.32&	  157.59&	157.31&	1&	0&	1&	5&	0&	5&	15&	0&	15&	0.65&	10&	2&	0.99\\
296&	19841120&	810&	10&	0.007&	19.91&	2.2&	  144.41&	144.32&	1&	1&	2&	6&	1&	7&	16&	11&	27&	0.65&	18&	3&	0.99\\
297&	19841121&	945&	145&	0.073&	19.59&	2.08&	  131.23&	130.26&	1&	1&	2&	13&	2&	15&	23&	12&	35&	0.64&	22&	3&	1.01\\
298&	19841122&	900&	100&	0.042&	19.26	&    1.95&	  118.05&	117.5&	1&	2&	3&	11&	4&	15&	21&	24&	45&	0.64&	29&	3&	1.01\\
299&	19841123&	840&	40&	0.028&	18.93&	1.83&  104.87&	104.5	&    1&	2&	3&	7&	15&	22&	17&	35&	52&	0.64&	33&	3&	1.01\\
300&	19841124&	940&	140&	0.07&	     18.59	&    1.71&   91.69&	90.77	&    1&	2&	3&	6&	18&	24&	16&	38&	54&	0.64&	35&	3&	1.01\\
301&	19841125&	820&	20&	0.014&	18.25&    1.58&	  78.51&	78.33	&    1&	2&	3&	2&	34&	36&	12&	54&	66&	0.64&	42&	3&	1.01\\

\hline 
\end{tabular}}}
\label{tabcsv1}
\end{table}

\begin{table}[h]
\caption{CSV Table Handwritten Parts} 
\centering 
\setlength{\tabcolsep}{1.5mm}{
\begin{tabular}{ccccccccc} 
\hline\hline 
Date & Order &Latitude&Longitude&Type&$\alpha$&$\alpha^{'}$&r&N
\\ \hline 
19580107&43&21&49&C&0.4&0.4&70&4\\
19580107&44&16&48&C&1.4&0.7&68&5\\
19580107&45&-19&40&A&0.2&0.2&58&2\\
19580107&47&-22&33&C&0.8&0.7&52&5\\
19580107&48&4&14&C&0.6&0.5&24&4\\
19580107&50&-17&1&C&1&0.9&20&3\\
19580107&51&11&-2&F&9&6&22&64\\
19580107&52&-28&-38&C&1.2&1&60&4\\
19580107&53&-17&-40&D&3.6&2.3&55&9\\
19580107&54&-11&-56&E&10&4&70&21\\
19580107&55&-16&5&B&0.1&0.1&21&4\\
19580107&56&-38&-13&C&0.6&0.4&52&6\\
19580107&57&14&-31&A&0.1&0.1&50&2\\
19580107&58&26&-68&J&0.8&0.8&82&1\\
\hline 
\end{tabular}}
\label{tabcsv2}
\end{table}
\section{Accuracy Verifications of Extraction Parameter}
Due to the large number of sunspot drawings, YNAO alone has 15,752 sunspot drawings, and there are 1,051,422 records of information in rectangular and oval boxes that contained in images. In order to trust the original data and ensure the reliability of digital data, the following verification methods are used to check data, and finally carried out the accuracy test.
\subsection{Check the Accuracy of Values by Data Type}
Each sunspot drawing contains different types of sunspot data, and the length (number of digits) of different types of values are different, but the data format of the same type of data is uniform. For example, in Table~\ref{tabcsv1}, number, date of observation, Beijing time, UTC, gN, gS, gNS, fN, fS, fNS and so on are positive integers, but P, B$_{0}$, L$_{0}$ and so on are decimals. The numeric type of the data can be determined by verifying the number of decimal places in the data, and checking whether the number of decimal places is consistent with the length of the standard numerical values.
\subsection{Abnormal Value Elimination}
Most of the sunspot drawing recorded by YNAO are limited to a specific range, e.g., the year of observation is 1958-2015; the monthly range is from January to December; For the range of days, the longer month is from 1 to 31 and the shorter month is from 1 to 30; February is from 1 to 29/28 for leap/common; P angle varies from -26.31 to 26.31; B$_{0}$ varies from -7.23 to 7.23; seeing is an integer between 1 and 5.
Different types of data are sorted separately, and then verified to determine whether the data is valid and its value is within the normal range. If not, the data is then manually compared with the data recorded in the sunspot drawings of that day.
\subsection{The Correlation Verification}
Correlation verification consists of two parts, one of which is the conversion and comparison of relevant data in the printed part. The printed part of the sunspot drawings recorded by YNAO contains information such as Beijing time, UTC, gN, gS, gNS, fN, fS, fNS, \textit{etc.} There exist certain conversion regularity, e.g., the time difference between Beijing time and UTC is about 8 hours; The number of sunspots on the visible solar surface is the sum of the number of sunspots on the northern and southern hemispheres, namely, the sum of gN and gS is equal to gNS. This rule can also be applied to fN, fS, fNS, and RN, RS, RNS. In addition, there are the following correlations between the number of sunspots, the number of sunspot groups, and the number of Wolf:\\
RN =10gN+fN\\
RS= 10gS+ fS\\
RNS =10gNS+fNS\\
R = K (10g + f), here R=RNS, f=fNS\\
The other part is to compare between the same data recorded in the printed and handwritten parts. For example, the printed part recorded the total number of sunspot groups and sunspots observed on that day, while the handwritten part recorded the specific information of each sunspot group, including the number of sunspots in each sunspot group. In this way, the total number of sunspot group on a given day can be obtained by counting the number of sunspots groups recorded in the handwritten part. When the number of sunspots labeled in each sunspot group is added, the total number of sunspots on that day can be obtained. Finally, it is compared with the values of gNS and fNS, respectively. Also, the method can be used to confirm if there is a sunspot group that has not been identified in handwritten part.

\subsection{Verification of the Relevant Laws}
In 1861, British astronomer Richard Christopher Carrington discovered what we now know as Sp\"orer's law, or butterfly diagram (\citeauthor{1976gaa..book.....H}, \citeyear{1976gaa..book.....H};\citeauthor{1969Carrington}, \citeyear{1969Carrington}), which was investigated in detail by German astronomer Gustav Sp\"orer. During the solar cycle, sunspots are distributed within a $\pm$ 45$^\circ$ latitude range, of which most sunspots appear on both sides of the equator within latitudes of 15$^\circ$-20$^\circ$ that parallel to the equator, however there are very few appear on both sides of the equator's 8$^\circ$ range. At the beginning of the sunspot cycle, sunspots mostly appear on the surface within 30$^\circ$ to 45$^\circ$ latitude. As the cycle progresses, sunspots appear in lower and lower latitudes. When the average latitude equals 15$^\circ$, the sunspot number reaches its maximum. The average latitude of sunspots will continue to decrease, at the end of the cycle the sunspots appear in about 7$^\circ$. Then the new cycle of sunspots began to appear at the relatively high latitudes (\citeauthor{1995IrAJ...22..228P}, \citeyear{1995IrAJ...22..228P}). For the processed sunspots, the validation is checked as follows: if latitude of sunspot group located between -45$^\circ$ to 45$^\circ$, and if the above rules are followed. If not, proofread them manually.
At the same time, in the solar cycle, some other parameters of the same sunspot group, such as the coordinates of longitude and latitude, area and number of sunspots, also change with the formation and decay of the sunspot group. For example, sunspots generally appear in the eastern hemisphere and disappear in the west. In this way, the longitude variation of the sunspot group can be verified, then be checked whether it conforms to Sp\"orer's law in each sunspot cycle. The size and number of sunspots in an individual sunspot group are usually increased from smaller to larger and then they reverse, of which the rule can also be used to verified the accuracy of data.

\subsection{Data Test}
By random sampling, a total of 160 sunspot drawings ($\sim$ 1\%) were selected to test human recognition and input errors. The specific test scheme is as follows: Setting up a group of 10 members, then divided into five teams. Each team is equipped with two computers, in which one to identify the original image information and the other to check the digitized data. 160 images are selected randomly from the dataset of sunspot drawings, in this sample the sunspot related information was 918 (namely, the total lines of Table~\ref{tabcsv1}) and the valid data was 10,864. According to the test results, the number of data errors in the fixed region (printed parts in Table~\ref{tabcsv1}) was 6, the number of error in the relevant information of sunspots was 29 (handwritten parts as Table~\ref{tabcsv2}), and the total number of data errors was 35. In this 160 data samples, the error rate of fixed area is 3.75\% (6/160); while for a line with information about a cluster of sunspots the error rate is 3.15\% (29/918). According to the calculation of valid data, the error rate for this sample with 160 sunspot drawings was 0.32\% (35/10864). The error data is shown in Table~\ref{testresult}.
\begin{table}[h]
    \begin{center}
   
         \caption{Test Results}\label{testresult}
          \resizebox{1.\hsize}{!}{
           \begin{tabular}{|c|c|c|c|c|}
            \hline
            Error Type& Error Year& Number of Error&Sum Errors&Total Errors\\
            \hline
            \multirow{3}{*}{\makecell[{}{p{1cm}}]{B$_{0}$}}&19610630&1 & \multirow{3}{*}{\makecell[{}{p{1cm}}]{3}} & \multirow{6}{*}{\makecell[{}{p{1cm}}]{ }} \\
\cline{2-3}
            &19631105&1&&6\\
\cline{2-3}
            &19770803&1&&printed\\
 \cline{1-4}
            \multirow{3}{*}{\makecell[{}{p{1cm}}]{P}}&19691217&1 & \multirow{3}{*}{\makecell[{}{p{1cm}}]{3}} &  parts\\
\cline{2-3}
            &19750621&1&&\\
\cline{2-3}
            &19830528&1&&\\
    \hline
  \multirow{3}{*}{\makecell[{}{p{1cm}}]{$\alpha$}}&19591201&1 & \multirow{3}{*}{\makecell[{}{p{1cm}}]{3}} & \multirow{26}{*}{\makecell[{}{p{1cm}}]{ }} \\
\cline{2-3}
            &19681109&1&&29 \\
\cline{2-3}
            &19830228&1&& handwritten\\
 \cline{1-4}
   \multirow{2}{*}{\makecell[{}{p{1cm}}]{$\alpha^{'}$}}&19680331&1 & \multirow{3}{*}{\makecell[{}{p{1cm}}]{3}} &  parts\\
\cline{2-3}
            &19681109&2&&\\
 \cline{1-4}
  \multirow{9}{*}{\makecell[{}{p{1cm}}]{r}}&19680331&2 & \multirow{9}{*}{\makecell[{}{p{1cm}}]{11}} & \\
\cline{2-3}
            &19681109&1&&\\
 \cline{2-3}
            &19741010&1&&\\
  \cline{2-3}
            &19781208&1&&\\
   \cline{2-3}
            &19810128&2&&\\
    \cline{2-3}
            &19890707&1&&\\
     \cline{2-3}
            &19890624&2&&\\
     \cline{2-3}
            &19920225&2&&\\
 \cline{1-4}
  \multirow{4}{*}{\makecell[{}{p{1cm}}]{Lati.}}&19680331&1 & \multirow{4}{*}{\makecell[{}{p{1cm}}]{4}} & \\
\cline{2-3}
            &19681109&1&&\\
  \cline{2-3}
            &19810421&1&&\\
              \cline{2-3}
            &19810503&1&&\\
              \cline{2-3}
            &19931226&1&&\\
  \cline{1-4}
   \multirow{2}{*}{\makecell[{}{p{1cm}}]{Long.}}&19800611&1 & \multirow{2}{*}{\makecell[{}{p{1cm}}]{2}} & \\
\cline{2-3}
            &19890409&1&&\\
 \cline{1-4}
   \multirow{4}{*}{\makecell[{}{p{1cm}}]{Num}}&19730505&1 & \multirow{4}{*}{\makecell[{}{p{1cm}}]{4}} & \\
\cline{2-3}
            &19780411&1&&\\
            \cline{2-3}
            &19830528&1&&\\
            \cline{2-3}

            &20010517&1&&\\
  \cline{1-4}
 \multirow{2}{*}{\makecell[{}{p{1cm}}]{ Group Type}}&19890624&1 & \multirow{2}{*}{\makecell[{}{p{1cm}}]{2}} & \\
\cline{2-3}
             &20040909&1&&\\
            \hline
\end{tabular}}

\end{center}
\end{table}
\section{Data Summary and Analysis}
\begin{figure}
   \centerline{\includegraphics[width=1\textwidth,clip=]{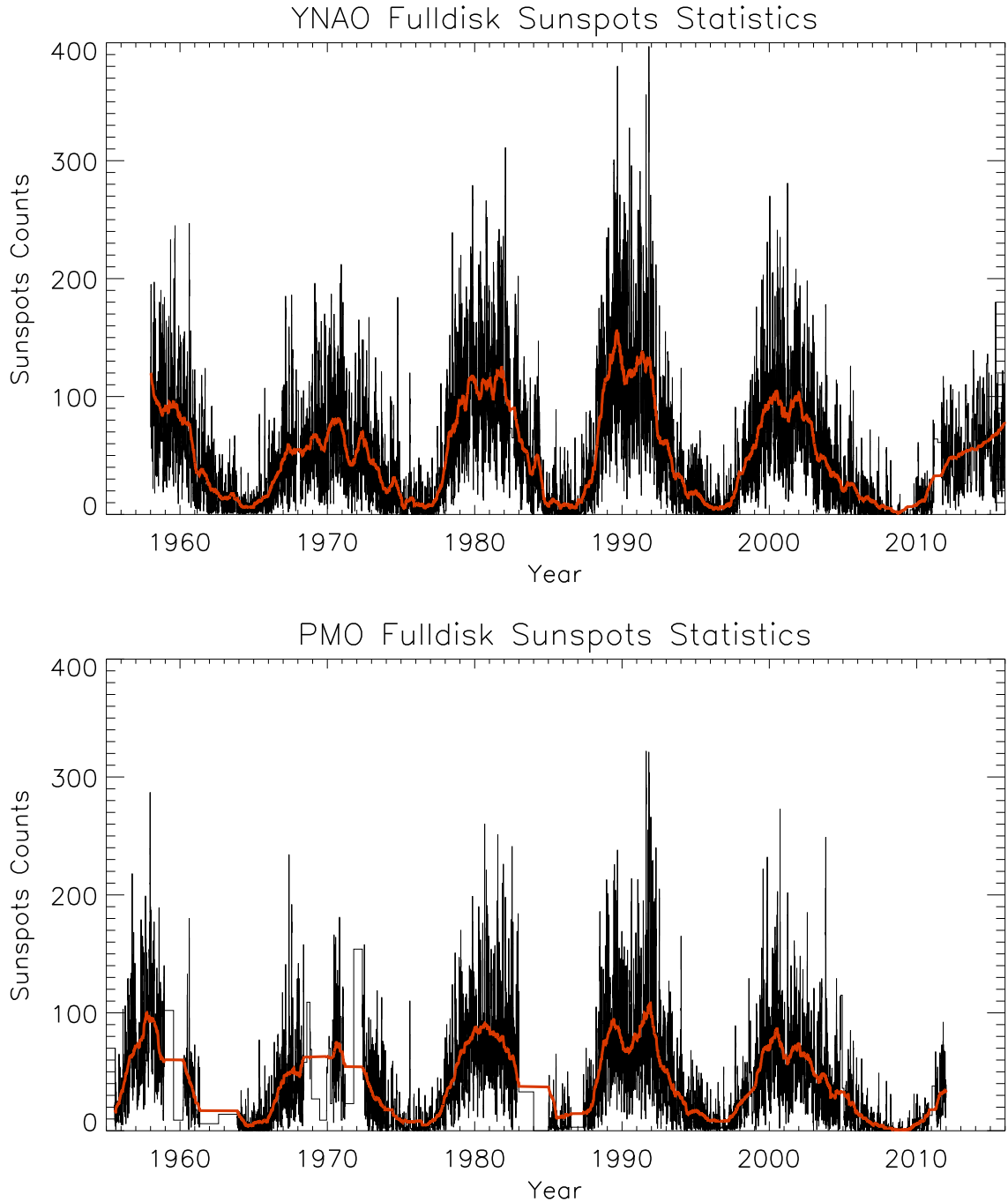}}
   \caption{Temporal evolution of sunspot number based on the digitized sunspot drawings observed by YNAO (upper panel) and PMO (bottom panel), here the sunspot numbers were directly observed by individual observing station.  This statisticals does not include the sunspots group information. The red solid curve is the 180 days smoothed curve superposed to the original data.} \label{ztytnum}
\end{figure}

\begin{figure}
   \centerline{\includegraphics[width=1\textwidth,clip=]{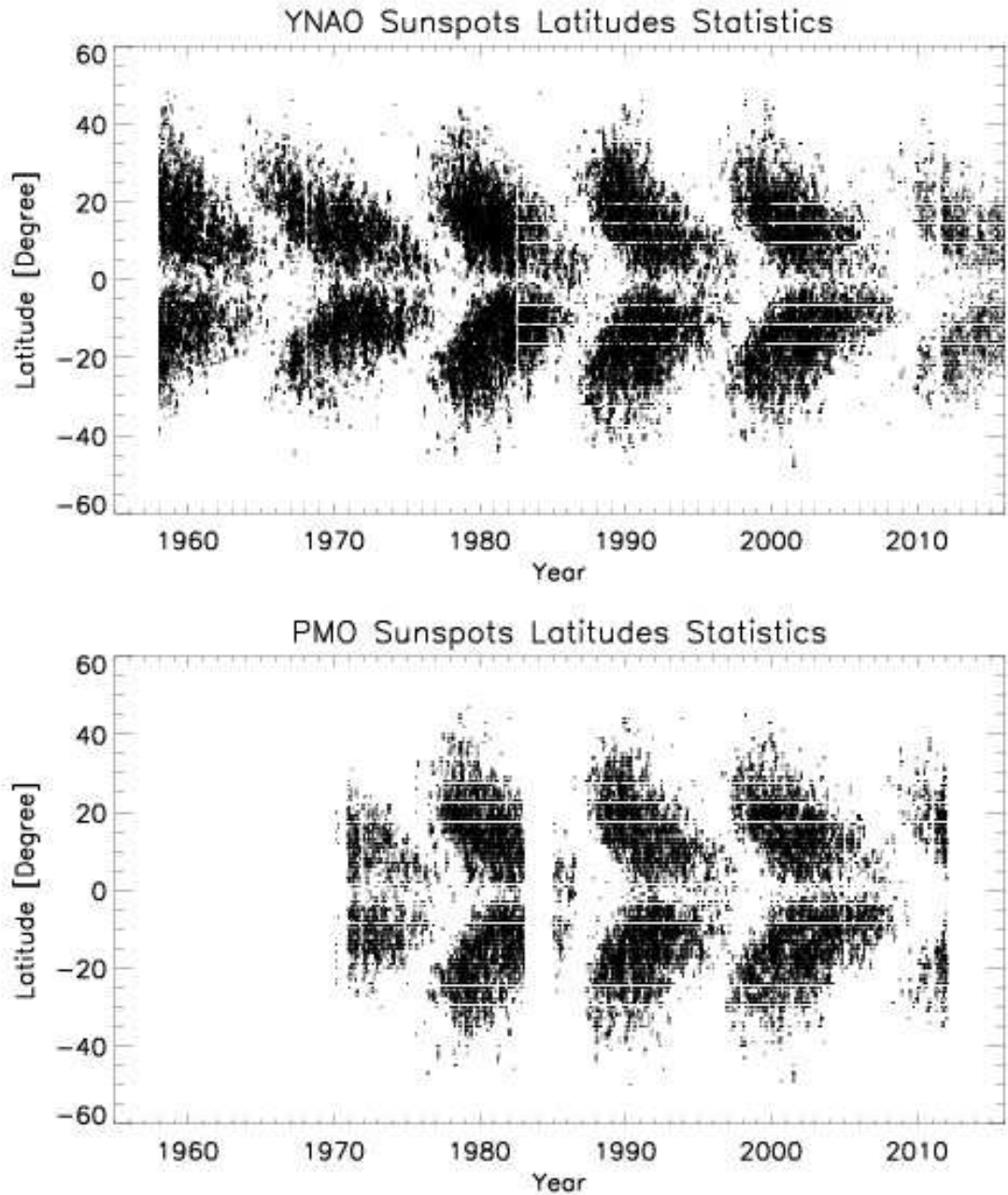}}
   \caption{Butterfly diagram according to heliographic latitudes recorded in the digitized sunspot drawings obtained by YNAO (upper panel) and PMO (bottom panel). The PMO sunspot drawing observers had the habit of only recording integer values for the sunspot location, resulting in 1$^\circ$ resolution. The YNAO observers tended to record float values before 1982, one more decimal place but mainly 0.5$^\circ$.} \label{ztytbf}
\end{figure}

Digitization, parameter extraction and quantitative statistics for decades of sunspot drawings in China are performed based on scanning mode of human--computer interaction and method of CNN handwritten character recognition. The results shown in Table~\ref{Digitizedresult} indicates the number of images are 47,073, for which the digitized results are stored in the format described in Section~\ref{Parameter Record Format}. Among them, the number of valid records included in the printed part is 47,073, and in handwritten part it is 235,489. Statistical distribution carried out on the digital data of PMO and YNO with a long observation time. Temporal evolutions of full-disk sunspot number based on the digitized sunspot drawings observed by YNAO (upper panel) and PMO (bottom panel) are shown in Figure~\ref{ztytnum}, where the X-coordinate is year and the Y-coordinate indicates the total number of sunspots directly observed by sunspot drawing telescope on a given day. The the red solid curve is the 180 days smoothed curve superposed to the original data. The sunspot drawings from YNAO and PMO together contributed the largest part to our data archives. In present, the pure full-disk sunspots counts, which do not include the group numbers, are accurate statistics for this data archives (the sunspot relative number R can be provided after the cross-calibration with the international sunspot relative number are finished). Figure~\ref{ztytbf} shows the periodic variation of sunspots obtained from the digital data of PMO and YNAO, where X-coordinate represents time (year) and the Y-coordinate mean latitude. This results provides the pattern of butterfly diagram, which shows the variation of sunspot latitudes in the northern and southern hemispheres over the past 58 years.
It is evident that the average solar cycle is 11 years, and the variation of the sunspot latitude is consistent with the Sp\"orer's law.. For example, at the beginning of the twentieth solar cycle (October 1964-June 1976), sunspots in the new cycle appeared to be at higher latitudes, and sunspots in the old cycle appeared at lower latitudes.

\begin{figure}
   \centerline{\includegraphics[width=1\textwidth,clip=]{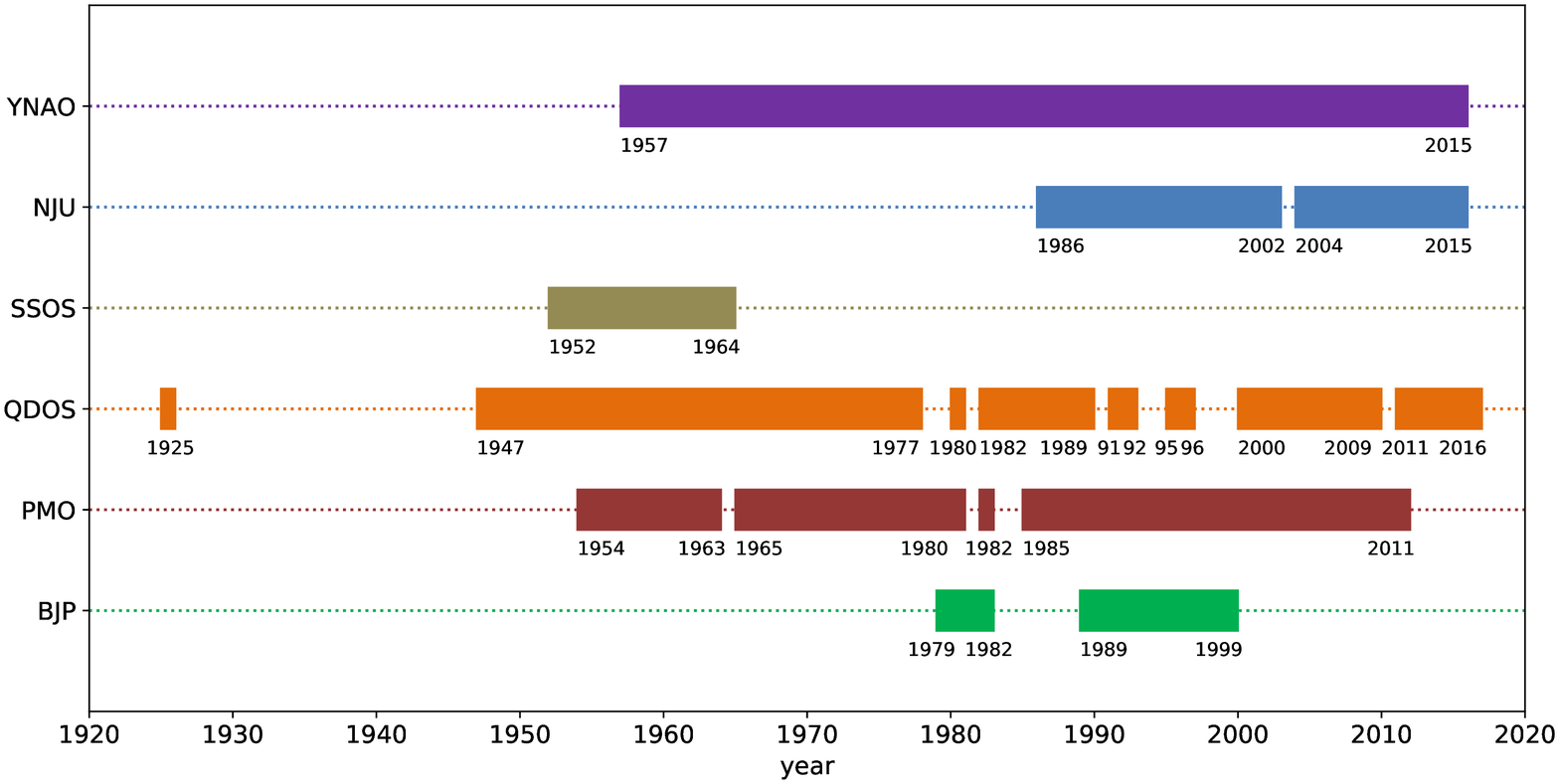}}
   \caption{Data continuity statistics of sunspot drawing at different observatories. The different colors of patche corresponds to individual observatory that was labeled on the Y-aaxis: violet, blue, darkkhaki, yellow, brown and green indicate YNAO, NJU, SSOS, QDOS, PMO and BJP. The observed years are plotted on the X-axis, which can show the time coverage of data that was obtained by different observatory. At the same time, the segmented observation periods for individual observation are labeled under corresponding color patche. } \label{conti}
\end{figure}
The statistics on the data continuity of each station are carried out, and the results are shown in Figure~\ref{conti}. Among them, PMO and YNAO have the longest observation time. Part of the monthly data is missing. We have counted the periods without sunspots drawing for more than seven days, and part of these results are given in Table~\ref{7day}. The YNAO has an interval of 100 times over seven days without images, while for PMO these are 283 times.
\begin{table}[h]
\caption{Statistical Results of Digitized Sunspot Drawing} 
\centering 
\setlength{\tabcolsep}{1.1mm}{
\begin{tabular}{|l|l|l|l|l|l|} 
\hline
Obs. & Years &Num. of &Printed Part&Handwritten Part&Total Num. of\\
&&Sunspot Drawing&(Records)&(Records)&Information\\ \hline 
YNAO&1957-2015&15752&15752&89262&1051422\\\hline
BJP&1979-1982&3764&3764&28647&170704\\
&1989-1999&&&&\\\hline
PMO&1954-1963&12508&12508&57570&626765\\
&1965-1980&&&&\\
&1982&&&&\\
&1985-2011&&&&\\\hline
QDOS&1925&12634&12634&48424&395047\\
&1947-1977&&&&\\
&1980&&&&\\
&1982-1989&&&&\\
&1991-1992&&&&\\
&1995-1996&&&&\\
&2000-2009&&&&\\
&2011-2016&&&&\\\hline
SSOS&1952-1964&2415&2415&11586&59471\\\hline
\end{tabular}}
\label{Digitizedresult}
\end{table}

\begin{table}[h]
\caption{Parts of No Observation Statistics for More than Seven Days} 
\setlength{\tabcolsep}{12mm}{
\begin{tabular}{|l|l|l|} 
\hline
Obs. & Years &Date\\\hline
YNAO&1958&0208-0214\\
&&1122-1201\\
\cline{2-3}
&1959&0221-0301\\
&&0524-0531\\
\cline{2-3}
&1960&0523-0613\\
&&0713-0721\\
\cline{2-3}
&1963&0101-0109\\
&&0608-0614\\
&&1016-1022\\
&&1226-0101\\
\hline
PMO&1956&0116-0123\\
&&0127-0204\\
&&0320-0402\\
&&0605-0612\\
&&0731-0807\\
&&0824-0831\\
&&0913-0925\\
\cline{2-3}
&1957&0107-0118\\
&&	0131-0213\\
&&	0420-0503\\
&&	0727-0807\\
&&	1123-1130\\
&&	1206-1215\\
\hline
\end{tabular}}
\label{7day}
\end{table}

Sunspot drawings originated from various regions of China are digitized, and the corresponding Software is developed to recognize and  extract the valuable parameters, then correctness verifications are also carried out at the same time. Additionally, according to the types and characteristics of data recorded in the image, a detailed data consolidation scheme and different data verification methods are developed. Based on this verification method, the accuracy of parameters are checked again. Finally, the database of sunspot drawings in China and their parameters was established. A preliminary analysis of the data shows that it is consistent with the significant pattern of sunspots. The work of digitizing Chinese sunspot drawings and extracting their parameters, not only makes Chinese historical sunspot observation permanently preserved, but also make the data to have continuity, integrity, complementarity and usability. These four aspects fill the gaps in the observation data of digital sunspots in this region, and provide more detailed and richer data for the study of long-term sunspots evolutions. The database is in the web site: http://sun.bao.ac.cn/SHDA\_data/.
At present, YNAO and QDOS continue to produce sunspot drawings, which can provide the subsequent continuous data for the ground-based observation of sunspots for the Chinese timezone.
The quality assessment of sunspot drawing in China is under way and corresponding results will be published. A web-based English version query interface will be released soon (database is already accessible by the web site: http://sun.bao.ac.cn/SHDA\_data/). The next step, we will pay more attention to these extracted parameters more accurately on the basis of combining original sunspot drawings and produced newer version parameters set.

\section{Data Policy}
The images and data from the Archive of Chinese Historical Sunspot Drawings (ACHSD) can be freely downloaded as public data. However, any public use, web based or paper publication of those data must include an explicit credit to the source: (ACHSD data/image, National Astronomical Observatories, CAS, Beijing ,China)

\acknowledgments
We thank the reviewer for valuable suggestions and constructive criticism, which improved the clarity of the article. The work was funded by National Science Foundation of China Grant Nos: u1531247 and 1142790111427901, the 13th Five-year InformatizationPlan of Chinese Academy of Sciences, Grant No. XXH13505-04 and the special foundation work of the ministry of science and technology of the of China Grant No: 2014fy120300. We would like to thank to the predecessors who have been engaged in the observation of sunspot in China over the past decades.


\end{document}